\documentclass[prb,showpacs,twocolumn,aps,superscriptaddress,a4paper]{revtex4-1}
\usepackage{color}
\usepackage{dcolumn}
\usepackage{amsmath}
\usepackage{graphicx}
\usepackage{latexsym}
\usepackage{amsfonts}
\usepackage{amssymb}
\usepackage{color}
\DeclareGraphicsExtensions{.pdf,.gif,.jpg}

\newcommand{\be}{\begin{equation}}
\newcommand{\ee}{\end{equation}}
\newcommand{\bea}{\begin{eqnarray}}
\newcommand{\eea}{\end{eqnarray}}

\def\a{\alpha}
\def\b{\beta}
\def\g{\gamma}
\def\G{\Gamma}
\def\d{\delta}
\def\D{\Delta}
\def\e{\epsilon}

\def\th{\theta}

\def\l{\lambda}

\def\m{\mu}

\def\p{\pi}

\def\r{\rho}
\def\s{\sigma}
\def\S{\Sigma}
\def\t{\tau}
\def\f{\phi}
\def\vf{\varphi}

\def\w{\omega}

\def\callT{\mbox{$\mathcal{T}$}}

\def\ra{\rightarrow}

\def\iif{\infty}
\def\bra{\langle}
\def\ket{\rangle}

\def\Tr{{\rm Tr}}

\def\Im{{\rm Im}}

\def\1op{\hat{\mathbbm{1}}}
\def\nn{\nonumber}

\begin{document}

\title{Charge dynamics in molecular junctions: Nonequilibrium 
Green's Function approach made fast}

\author{S. Latini}
\affiliation{Dipartimento di Fisica, Universit\`{a} di Roma Tor Vergata,
Via della Ricerca Scientifica 1, 00133 Rome, Italy}
\author{E. Perfetto}
\affiliation{Dipartimento di Fisica, Universit\`{a} di Roma Tor Vergata,
Via della Ricerca Scientifica 1, 00133 Rome, Italy}
\author{A.-M. Uimonen}
\affiliation{Department of Physics, Nanoscience Center, FIN 40014, University of Jyv\"askyl\"a,
Jyv\"askyl\"a, Finland}
\author{R. van Leeuwen}
\affiliation{Department of Physics, Nanoscience Center, FIN 40014, University of Jyv\"askyl\"a,
Jyv\"askyl\"a, Finland}
\affiliation{European Theoretical Spectroscopy Facility (ETSF)}
\author{G. Stefanucci}
\affiliation{Dipartimento di Fisica, Universit\`{a} di Roma Tor Vergata,
Via della Ricerca Scientifica 1, 00133 Rome, Italy}
\affiliation{INFN, Laboratori Nazionali di Frascati, Via E. Fermi 40, 00044 Frascati, 
Italy}
\affiliation{European Theoretical Spectroscopy Facility (ETSF)}

\begin{abstract}
Real-time Green's function simulations of molecular junctions 
(open quantum systems) are typically performed by solving the 
Kadanoff-Baym equations (KBE). The KBE, however, impose a serious 
limitation on the maximum propagation time due to the large memory 
storage needed.  In this work we propose a simplified Green's 
function approach based 
on the Generalized Kadanoff-Baym Ansatz (GKBA) to overcome the KBE 
limitation on time, significantly speed up the calculations, and yet stay 
close to the KBE results. This is achieved through a twofold advance: 
first we show how to make the GKBA work in open systems and then 
construct a suitable quasi-particle propagator that includes
correlation effects in a diagrammatic fashion. We also provide 
evidence that our GKBA scheme, although already in good 
agreement with the KBE approach, can be further improved without 
increasing the computational cost.
\end{abstract}


\maketitle

\section{Introduction}
Charge transfer through nanoscale interfaces is an ubiquitous 
dynamical process in molecular electronics, photovoltaics, 
electroluminiscence and transient spectroscopy, to mention a few emerging fields of 
research.\cite{zhu.2004,n.book}  The complexity of the 
molecules (or molecular aggregate) and of the contacts to a 
source/drain electrode, as well as the simultaneous interplay of Coulomb repulsion and 
vibrational effects make these research fields an interdisciplinary 
topic where physics, chemistry and engineering meet. Reliable theoretical predictions 
require an accurate 
description of the nuclear degrees of freedom, a careful selection of 
the electronic basis functions and a proper  treatment 
of correlation effects. 

Among the {\em ab 
initio} methods, Density Functional Theory\cite{dg.book,py.book} (DFT) and its Time 
Dependent extension\cite{tddft.book,u.book} (TDDFT) stand out for the advantageous scaling of 
the  computational cost with increasing the system size and the 
propagation time. However, as for any other method, a (TD)DFT 
implementation is based on 
some approximation and, at present, the available approximations are 
inadequate to capture  correlation effects 
like the Coulomb blockade\cite{CB2,hj.book,cs.book} or
the polarization-induced renormalization of the molecular 
levels.\cite{Neaton,kr1.2009,kr2.2009,Kaasbjerg1,mtksvl.2012} 
These effects are particularly important in a donor-acceptor complex, in a molecular junction in 
the weak-coupling regime and more generally when
the transition rate for an electron to move from one atom to another is 
small. Many-body approaches based on 
Nonequilibrium Green's Functions\cite{svl.book} (NEGF) offer a 
promising alternative as the relevant scattering processes 
to describe the aforementioned effects
can be incorporated either
through  a proper selection of Feynman diagrams or through a decoupling 
scheme for the higher order Green's functions. Real-time simulations 
within the NEGF are performed by solving the Kadanoff-Baym 
equations\cite{svl.book,kb.book,kb.2000,dvl.2007} (KBE), which are a 
set of coupled nonlinear integro-differential equations 
for the one-particle Green's function.
Unfortunately, the price to pay in solving the KBE 
is that the computational time 
scales cubically with the  propagation time (given the self-energy) whereas in TDDFT the 
scaling is linear
(given the exchange-correlation potential). 

In the mid eighties Lipavsky {\em et al.}\cite{lsv.1986} proposed an approximation to 
scale down the computational time (from cubic to quadratic) of the KBE. This 
approximation is known as the Generalized Kadanoff-Baym Ansatz (GKBA) 
and has been successfully applied to strongly interacting nuclear 
matter,\cite{k.1996} electron 
plasma,\cite{bksbkk.1996,bkbk.1997,kbbk.1998,bsh.1999}, carrier dynamics  
of semiconductors,\cite{h.1992,hj.book,m.2013}, optical 
absorption spectra,\cite{pphs.2011}  quasi-particle
spectra,\cite{ppsh.2009} and more recently 
excited Hubbard clusters.\cite{hb.2013,bhb.2013} 
In all these cases the system is either a bulk periodic system or a 
finite system. It is currently unknown
how the GKBA performs for 
nanostructures 
chemically bonded to or adsorbed on a surface (open system). 
In fact, in open systems a number of 
issues have to be addressed before a GKBA 
calculation can be carried out. For instance the GKBA remains an 
approximation even in a noninteracting (or mean-field) treatment whereas 
in closed systems  it is exact.
Furthermore the performance of the GKBA strongly depends 
on the quality of the quasi-particle propagator and, as we shall see, 
in open systems 
the available approximations perform rather poorly.

This work contains a thorough study of the GKBA in open 
systems. In Section \ref{gkbaopensec} we derive the fundamental equations and 
present a few exact properties. Here the discussion is mainly 
focussed on noninteracting and mean-field electrons. Important aspects of the 
GKBA like the construction of a mean-field propagator as well as issues 
related to relaxation and local thermalization are analyzed and 
addressed. This preliminary investigation is particularly relevant since, as 
previously mentioned, the GKBA is an approximation already at the 
mean-field level. In the correlated case the 
GKBA simulations using a mean-field propagator are far 
off the KBE results. In Section \ref{corrpropsec} we propose a couple 
of correlated propagators to remedy this deficiency. Our propagators have the merit of scaling 
quadratically with the propagation time and hence the computational 
gain of the GKBA is maintained. The different GKBA schemes are 
compared with the full KBE approach in Section \ref{results}.
 We consider two  systems, a molecular junction under applied bias and a 
donor-acceptor complex under illumination, and calculate local 
currents and densities. Both systems constitute a severe test for the 
GKBA as the inclusion of correlations changes dramatically the 
mean-field picture. The important message emerging from this study 
is that one of the proposed GKBA schemes is in 
fairly, sometimes extremely, good agreement with the KBE approach.  
We also provide numerical evidence that the GKBA scheme can
be further improved at the same computational cost.
In conclusion, time-dependent simulations of open systems within the NEGF framework 
can be made much faster. 

\section{GKBA in open systems}
\label{gkbaopensec}

In this Section we briefly review the KBE for open systems and 
discuss in detail the simplifications brought about by the GKBA. 
The most general Hamiltonian which describes a molecular junction in 
contact with $M$ electronic reservoirs has the form
\be
\hat{H}=\sum_{\a=1}^{M}\hat{H}_{\a}+\hat{H}_{J}+\hat{H}_{T}.
\label{htot}
\ee
In Eq. (\ref{htot}) the Hamiltonian of the  $\a$ reservoir reads
\be
\hat{H}_{\a}=\sum_{k\a\s}\e_{k\a}\hat{d}_{k\a\s}^{\dag}\hat{d}_{k\a\s}
\ee
with $\hat{d}_{k\a\s}$ the annihilation 
operator for electrons of spin $\s$ and energy $\e_{k\a}$.  
The Hamiltonian of the molecular junction is expressed in terms of the 
operators $\hat{d}_{i\s}$ for electrons of spin $\s$ in the $i$-th 
localized molecular orbital 
\be
\hat{H}_{J}=\sum_{\substack{ij\\ \s}}h_{ij}\hat{d}_{i\s}^{\dag}\hat{d}_{j\s}+
\frac{1}{2}\sum_{\substack{ijmn\\ \s\s'}}
v_{ijmn}\hat{d}_{i\s}^{\dag}\hat{d}_{j\s'}^{\dag}\hat{d}_{m\s'}\hat{d}_{n\s}
\label{hamnano}
\ee
where $h_{ij}$ are the one-electron matrix elements of the one-body 
part (kinetic plus potential energy) and $v_{ijmn}$ are the 
two-electron Coulomb integrals. The last term in Eq. (\ref{htot}) is 
the tunneling Hamiltonian between the 
different subsystems and reads
\be
\hat{H}_{T}=\sum_{k\a\s}\sum_{i}\left(T_{k\a,i}\hat{d}_{k\a\s}^{\dag}\hat{d}_{i\s}+{\rm H.c.}
\right)
\ee
with $T_{k\a,i}$ the tunneling amplitude between the $i$-th state of 
the molecular junction and the $k$ state of the $\a$ reservoir.

Initially, say at time $t=0$, the system is in equilibrium at inverse 
temperature $\b$ and chemical potential $\m$. We assume that this 
equilibrium state can be reached starting from  the uncontacted 
($T_{k\a,i}=0$) and noninteracting ($v_{ijmn}=0$) system in the 
remote past, $t=-\iif$, and then propagating forward in time with the full 
interacting and contacted Hamiltonian until $t=0$. This  
amounts to assume that initial-correlation and memory effects are 
washed out. In our experience this assumption is always 
verified.\cite{mssvl.2008,psc.2010}  At time $t=0$ 
the system is driven out of equilibrium  by external electromagnetic 
fields, $\e_{k\a}\ra \e_{k\a}+V_{\a}(t)$ and $h_{ij}\ra h_{ij}(t)$. 
We are interested in monitoring the evolution of the electronic 
degrees of freedom through the calculation of observable quantities 
like, e.g., the local occupation and current.

\subsection{Green's function and KBE}

\begin{figure}[tbp]
\includegraphics[width=0.47\textwidth]{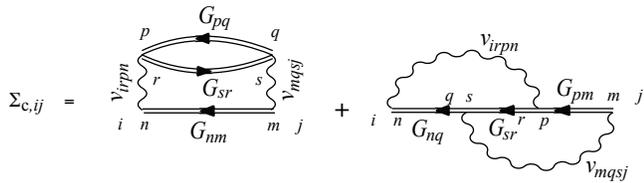}
\caption{Diagrams for the 2B correlation self-energy.}
\label{Sigma_2B}
\end{figure}

The building block of any diagrammatic many-body approach is the 
Green's function defined according to\cite{svl.book}
\be
G_{ij}(z,z')=\frac{1}{i}\bra\callT\left\{\hat{d}_{i\s,H}(z)\hat{d}^{\dag}_{j\s,H}(z')\right\}\ket.
\ee
In this definition the symbol ``$\bra \ldots \ket$'' denotes a gran-canonical 
average, and $\callT$ is the contour ordering acting on operators in 
the Heisenberg picture. The Green's function has arguments $z$ and 
$z'$ on the contour $\g$ going from $-\iif$ to 
$\iif$ (forward branch) and back from $\iif$ to 
$-\iif$ (backward branch). On this contour $G_{ij}$
satisfies the equations of 
motion\cite{mssvl.2009} (in matrix form)
\be
\left[
i\frac{d}{dz}-h_{\rm HF}(z)\right]G(z,z')=\d(z,z')+\!
\int_{\g} d\bar{z}\,\S(z,\bar{z})G(\bar{z},z')
\label{eomG}
\ee
and its adjoint.
Let us describe the various quantities in this equation.
The Hartree-Fock (HF) single-particle Hamiltonian is the sum of $h$ and 
the HF potential
\be
h_{{\rm 
HF},ij}=h_{ij}+\sum_{mn}\left(2v_{imnj}\r_{nm}-v_{imjn}\r_{nm}\right)
\ee
where 
\be
\r_{nm}(z)\equiv -iG_{nm}(z,z^{+})
\ee
is the time-dependent single-particle 
density matrix. The kernel $\S=\S_{\rm em}+\S_{\rm c}$ is the sum of 
the so called embedding self-energy and the correlation self-energy. 
The former can be calculated directly from the parameters of the 
Hamiltonian and reads
\be
\S_{{\rm em},ij}(z,z')=\sum_{k\a}T_{i,k\a}g_{k\a}(z,z')T_{k\a,j}
\ee
where 
\be
g_{k\a}(z,z')=\frac{1}{i}\left[\th(z,z')\bar{f}(\e_{k\a})-\th(z',z)f(\e_{k\a})\right]
e^{-i\f_{k\a}(z,z')}
\label{gka}
\ee
is the Green's function of the disconnected $\a$ reservoir. In Eq. 
(\ref{gka})  $f(\e)=1/(e^{\b(\e-\m)}+1)$ is the Fermi 
function, $\bar{f}(\e)=1-f(\e)$ and the phase $\f_{k\a}(z,z')=
\int_{z'}^{z}d\bar{z}(\e_{k\a}+V_{\a}(\bar{z}))$.
The expression of the correlation self-energy depends on the 
choice of diagrams that we decide to include. In this work we 
consider the second-Born (2B) approximation  which has been shown 
to produce results very close to those of the GW 
approximation,\cite{mssvl.2008} and to those of
numerically exact techniques in model systems.\cite{ukssklg.2011}
The 2B self-energy is given by the 
sum of the lowest order bubble 
diagram plus the second-order exchange diagram, see Fig. 
\ref{Sigma_2B},\cite{nota2} 
\bea
\S_{{\rm c},ij}(z,z')&=&
\sum_{nmpqrs}v_{irpn}v_{mqsj}
\nn\\
&\times&
\left[2G_{nm}(z,z')G_{pq}(z,z')G_{sr}(z',z)\right.
\nn\\
&-&\left.G_{nq}(z,z')G_{sr}(z',z)G_{pm}(z,z') \right]
\label{2bse}
\eea

To solve Eq. (\ref{eomG}) we convert it into a set of coupled 
equations, known as the KBE, for real time 
(as opposed to contour time) quantities. This 
is done by letting $z$ to vary on the forward (backward) branch and 
$z'$ to vary on the backward  (forward) branch of the contour $\g$. 
Using the Langreth rules\cite{svl.book,langreth} to convert contour-time convolutions into 
real-time convolutions we find (in matrix form)
\bea
\left[
i\frac{d}{dt}-h_{\rm HF}(t)\right]G^{<}(t,t')&=&I^{<}(t,t')
\label{kb1}
\\
G^{>}(t,t')\left[
-i\frac{\overleftarrow{d}}{dt'}-h_{\rm HF}(t')\right]&=&I^{>}(t,t')
\label{kb2}
\eea
with collision integrals
\be
I^{<}(t,t')=
\int_{-\iif}^{\iif} d\bar{t}\left[\S^{<}(t,\bar{t})G^{\rm A}(\bar{t},t')+
\S^{\rm R}(t,\bar{t})G^{<}(\bar{t},t')\right],
\label{collint<}
\ee
\be
I^{>}(t,t')=
\int_{-\iif}^{\iif} d\bar{t}\left[G^{>}(t,\bar{t})\S^{\rm A}(\bar{t},t')+
G^{\rm R}(t,\bar{t})\S^{>}(\bar{t},t')\right].
\label{collint>}
\ee
Here the superscripts ``$>,\;<,\;{\rm R,\; A}$'' refer to the lesser, 
greater, retarded and advanced Keldysh components. Equations 
(\ref{kb1},\ref{kb2}) are solved by a time stepping technique, 
starting from a value $G^{\lessgtr}(t_{\rm in},t_{\rm in})$  at some 
initial time $t_{\rm in}<0$  and then evolving along the directions $t$ 
and $t'$ until a maximum propagation time $t_{\rm max}$. The time 
$t_{\rm in}$ is chosen remotely enough in the past in order to have 
full relaxation at $t=0$, time at which the external fields are 
switched on.\cite{note} As 
$I^{\lessgtr}(t,t')$ in Eqs. (\ref{collint<},\ref{collint>}) involves integrals between $t_{\rm 
in}$ (the self-energy vanishes for times smaller than $t_{\rm in}$ 
since the system is initially uncontacted and noninteracting) and 
either $t$ or $t'$, the numerical effort in solving the KBE 
scales like $t_{\rm max}^{3}$.

\subsection{GKBA}

The GKBA allows us to reduce drastically the computational time. The 
basic idea consists in obtaining a closed equation for the equal time 
$G^{<}$ from which to calculate the time-dependent averages of all one-body 
observables like, e.g., density, current, dipole moment, etc. 
The GKBA is therefore an ansatz for the density matrix  
$\r(t)=-iG^{<}(t,t)$, not for the spectral function which has 
to be approximated separately, see below.

The exact equation for $\r(t)$ follows from the difference between Eq. 
(\ref{kb1}) and its adjoint, and reads 
\be
\frac{d}{dt}\r(t)+i\left[h_{\rm 
HF}(t),\r(t)\right]=-\left(I^{<}(t,t)+{\rm H.c.}\right).
\label{eomrho}
\ee
This is not a closed equation for $\r$ as the collision integral 
contains the off-diagonal (in time) $G^{\lessgtr}$. To close Eq. (\ref{eomrho}) we make the 
GKBA\cite{lsv.1986} 
\bea
G^{<}(t,t')&=&iG^{\rm 
R}(t,t')G^{<}(t',t')-iG^{<}(t,t)G^{\rm A}(t,t')
\nn\\
&=&-G^{\rm R}(t,t')\r(t')+\r(t)G^{\rm A}(t,t')
\label{gkba<}
\eea
and similarly
\be
G^{>}(t,t')=G^{\rm R}(t,t')\bar{\r}(t')-\bar{\r}(t)G^{\rm A}(t,t')
\label{gkba>}
\ee
where $\bar{\r}(t)=1-\r(t)=iG^{>}(t,t)$. However, the GKBA 
alone is not enough to close Eq. (\ref{eomrho}) since the 
quasi-particle propagator $G^{\rm R}$ 
(and hence $G^{\rm A}=[G^{\rm R}]^{\dag}$), or equivalently the 
spectral function, 
remains unspecified. The possibility of using the 
GKBA in open systems strongly relies on the choice of $G^{\rm R}$. This is an 
important point which we thoroughly address in the next Section. For 
the time being we observe that the numerical effort in solving Eq. 
(\ref{eomrho}) scales like $t_{\rm max}^{2}$ provided that the 
calculation of $G^{\rm R}$ does not scale faster.\cite{nota3} 

\subsubsection{Exact properties}

Among the properties of the GKBA we mention the fulfillment of the 
relation $G^{\rm R}-G^{\rm A}=G^{>}-G^{<}$ for any choice of $G^{\rm 
R}$, and the fact that Eqs. (\ref{gkba<},\ref{gkba>}) become an identity
in the limit $t\ra t'$ since $G^{\rm R}(t^{+},t)=-i$. Another 
valuable feature (in systems out of equilibrium) is that the GKBA 
preserves the continuity equation. There is, 
however, an even more important property from which 
the physical contents of the GKBA become evident. In closed systems ($\S_{\rm 
em}=0$) and for HF electrons ($\S_{\rm c}=0$) the collision integrals 
vanish and Eqs. 
(\ref{gkba<},\ref{gkba>}) are the solution of 
Eqs. (\ref{kb1},\ref{kb2})
provided that  $G^{\rm R}$ is the HF 
propagator\cite{{hj.book},svl.book}
\be
G^{\rm R}(t,t')=-i\th(t-t')\,T\,e^{-i\int_{t'}^{t}d\bar{t}\,h_{\rm 
HF}(\bar{t})}
\label{hfgr}
\ee
where $T$ is the time-ordering operator. 
Therefore, the more the quasi-particle 
picture is valid the more the GKBA is accurate. A more exhaustive 
discussion on the range of applicability of the GKBA in closed systems can be found in 
Refs. \onlinecite{lsv.1986,bk.1996}. 

In open systems the GKBA is 
not the solution of the HF equations
since $\S_{\rm em}\neq 0$ and hence the collision integral 
is  nonvanishing. The reliability of the GKBA in open systems needs 
to be investigated already at the HF level. In HF  
the collision integrals are evaluated with $\S=\S_{\rm em}$ and 
$G^{\rm R}$ being the solution of
\be
\left[
i\frac{d}{dt}-h_{\rm HF}(t)\right]\!G^{\rm R}(t,t')=
\d(t,t')+\int \!\! d\bar{t}\,\S^{\rm R}_{\rm em}(t,\bar{t})G^{\rm 
R}(\bar{t},t').
\label{eqgr}
\ee
In HF-GKBA the 
collision integrals are evaluated with 
$\S=\S_{\rm em}$, $G^{<}(\bar{t},t')=\r(\bar{t})G^{\rm 
A}(\bar{t},t')$ and $G^{\rm A}=[G^{\rm R}]^{\dag}$ some suitable propagator. If we 
calculate $G^{\rm R}$ from Eq. (\ref{eqgr})  then the 
numerical advantage of the GKBA is lost since the computational cost 
of solving this equation scales like $t_{\rm max}^{3}$. Thus the 
questions are:
can a ``computationally cheap'' 
propagator be constructed for open systems? If so, 
how accurate is the solution of the HF-GKBA equation?

To answer these questions we consider a Wide Band Limit (WBL) 
embedding self-energy $\S^{\rm R}_{\rm em}(t,t')=-(i/2)\G\d(t-t')$ 
where $\G$ is a positive-semidefinite self-adjoint matrix.
In this case the solution of Eq. (\ref{eqgr}) is
\be
G^{\rm R}(t,t')=-i\th(t-t')\,T\,e^{-i\int_{t'}^{t}d\bar{t}\,(h_{\rm 
HF}(\bar{t})-i\G/2)}
\label{hfgrwbl}
\ee
which has the same mathematical structure of Eq. 
(\ref{hfgr}). In particular it has the group property 
\be
G^{\rm R}(t+\d,t')=iG^{\rm R}(t+\d,t)G^{\rm R}(t,t')
\label{group}
\ee
and hence the number of operations to calculate $G^{\rm R}$ for all 
$t<t_{\rm max}$ and $t'<t$ scales like $t_{\rm max}^{2}$. 
The HF collision integral reads
\be
I^{<}(t,t)=
\int_{-\iif}^{\iif} d\bar{t}\,\S_{\rm em}^{<}(t,\bar{t})G^{\rm A}(\bar{t},t)
-\frac{i}{2}\G G^{<}(t,t),
\label{collinthf}
\ee
whereas the HF-GKBA  collision integral reads 
\be
I^{<}(t,t)=
\int_{-\iif}^{\iif} d\bar{t}\,\S_{\rm em}^{<}(t,\bar{t})G^{\rm A}(\bar{t},t)
-\frac{i}{2}\G \r(t)G^{\rm A}(t^{-},t).
\label{collinthfgkba}
\ee
If in Eq. (\ref{collinthfgkba}) we use for $G^{\rm A}=[G^{\rm R}]^{\dag}$ 
the HF result in Eq. (\ref{hfgrwbl}) then the collision integrals 
are identical since $G^{\rm A}(t^{-},t)=i$ and $i\r(t)=G^{<}(t,t)$. 
We conclude that the $G^{<}(t,t)$ that solves the HF and HF-GKBA equations is the {\em same}
provided that we use the same $G^{\rm R}$ of Eq. 
(\ref{hfgrwbl}). This observation contains useful hints on 
how to approximate the quasi-particle propagator of open systems without 
paying a too high computational price. 
We emphasize that the locality in time of the retarded embedding 
self-energy and of the HF self-energy $\S_{\rm 
HF}(z,z')=\d(z,z')[h_{\rm HF}(z)-h(z)]$ are distinct and should not be lumped together.
The former is purely imaginary and hence $\S^{<}_{\rm em}\neq 0$ 
whereas the latter is purely real and hence $\S^{<}_{\rm HF}=0$. 
Alternatively we can say that $\S_{\rm HF}$ is local on the contour whereas 
$\S_{\rm em}$ is not. This 
is a crucial difference: in closed systems the 
off-diagonal HF-GKBA $G^{<}(t,t')$ is the same as the  HF $G^{<}(t,t')$ whereas in open systems it 
remains an approximation even for a WBL embedding self-energy. Only
the diagonal HF and HF-GKBA $G^{<}(t,t)$ are identical in this case.

\begin{figure}[tbp]
\includegraphics[width=0.47\textwidth]{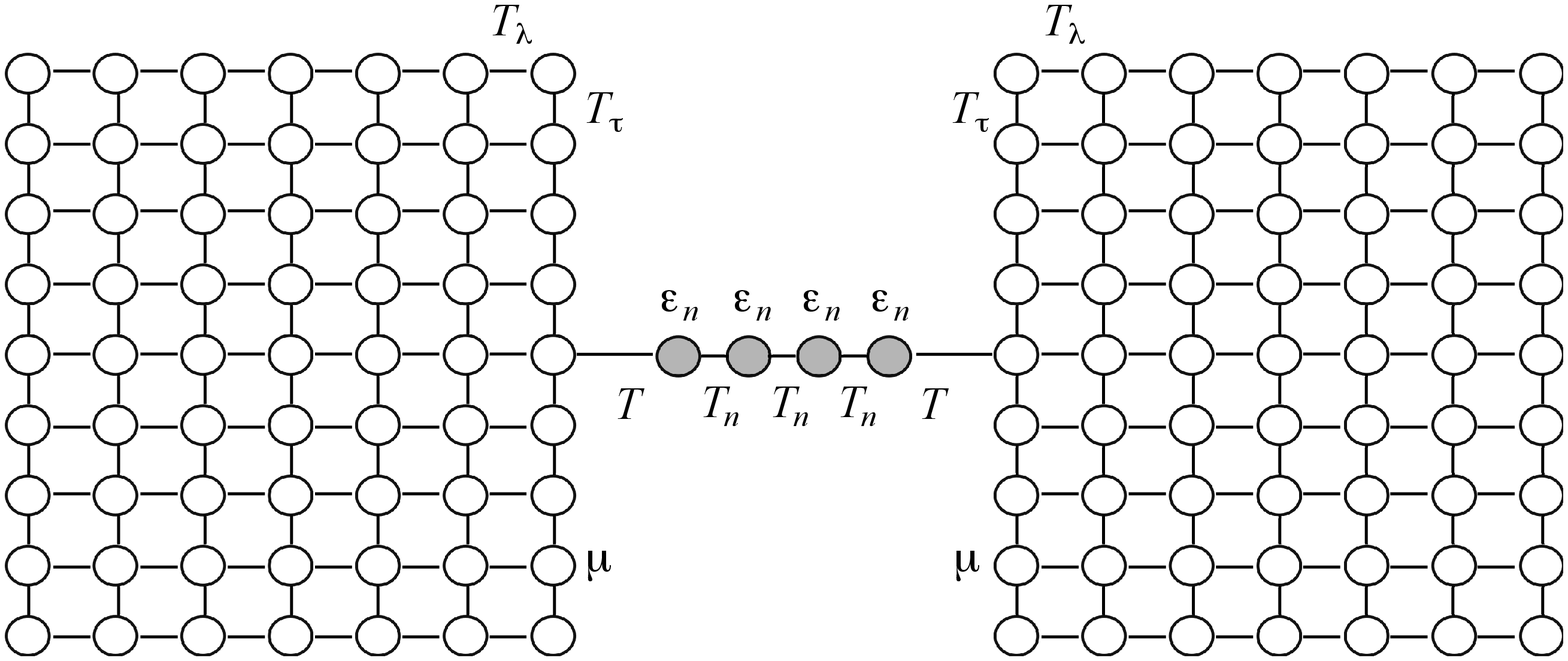}
\caption{Example of an open system as described in the main 
text with $N_{\t}=9$ transverse channels and a 
chain of four sites. }
\label{dleads}
\end{figure}

\subsubsection{An approximate propagator for mean-field electrons}
\label{hfappgr}

In most physical situations the removal and addition energies
relevant to describe the electron dynamics of the 
molecular junction after the application of a voltage difference or a laser 
pulse are well inside the continuum spectrum of the 
reservoirs. It is therefore natural to study how well the GKBA 
equation performs when $G^{\rm R}$ is chosen as in Eq. 
(\ref{hfgrwbl}) with 
\be
\G=i\left[\S_{\rm em}^{\rm R}(\m)-\S_{\rm em}^{\rm A}(\m)\right].
\label{gamma}
\ee
In Eq. (\ref{gamma}) the quantity $\S_{\rm em}^{\rm R}(\m)$ is the 
Fourier transform of the equilibrium embedding self-energy evaluated 
at the chemical potential.
This choice of $\G$ is expected to yield accurate 
results whenever $\S_{\rm em}^{\rm R}(\w)$ depends weakly on 
$\w$ for 
frequencies around $\m$.
Let us address this issue numerically.
We consider a class of systems consisting of two reservoirs, 
$\a=L,R$, with 
$N_{\t}$ transverse channels and a nanostructure with a 
chain geometry, see Fig. \ref{dleads}. We use a tight-binding representation 
and characterize the Hamiltonian of the reservoirs by a 
transverse hopping $T_{\t}$ and a longitudinal hopping $T_{\l}$ between 
nearest neighboring sites, and an onsite energy $\e=\m$ (half-filled 
reservoirs). The 
molecular chain has matrix elements $h_{ij}=T_{n}$ between nearest neighboring 
sites $i$ and $j$ and $h_{ii}=\e_{n}$ on the diagonal. The left 
reservoir is contacted through its middle terminal site to the 
leftmost site of the chain while the right reservoir is contacted through its middle terminal site to the 
rightmost site of the chain. We denote by $T$ the corresponding matrix elements of the 
Hamiltonian.\cite{nota4}

\begin{figure}[tbp]
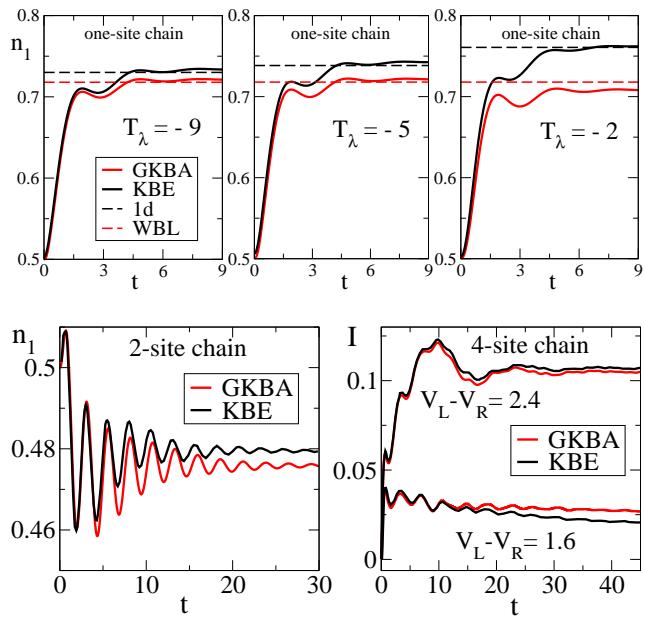

\includegraphics[width=0.47\textwidth]{fig3a.eps}
\includegraphics[width=0.47\textwidth]{fig3b.eps}
\caption{Top panels: density $n_{1}=\r_{11}$ of a one-site chain connected to leads 
with $N_{\t}=1$ after the sudden switch-on of a bias $V_{L}=2$
for different $T_{\l}=-9,-5,-2$. Bottom left 
panel: HF density of site 1 of a two-site chain connected to leads 
with $N_{\t}=1$
after the sudden switch-on of a bias $V_{L}=-V_{R}=1$. 
Bottom right panel: HF current at the right 
interface of a four-site chain connected to leads with $N_{\t}=9$
after the sudden switch-on of a bias $V_{L}=-V_{R}=0.8,1.2$.
}
\label{fig2}
\end{figure}
In Fig. \ref{fig2} we compare GKBA versus full KBE results 
for noninteracting and HF electrons. In all cases the Coulomb 
integrals $v_{ijmn}=\d_{in}\d_{jm}v_{ij}$.
The top panels refer to a system with 
$N_{\t}=1$ and a single-site chain driven out of equilibrium by a 
bias $V_{L}=2$ and $V_{R}=0$. The parameters (in arbitrary 
units) are $\m=\e_{n}=0$, 
$v_{ij}=0$, and $T=\sqrt{\g |T_{\l}|/2}$  with $\g=0.4$. From Eq. 
(\ref{gamma}) we find $\G=2\g$. The simulations have been performed at 
zero temperature for three different 
values of $T_{\l}=-9,-5,-2$, and are compared with exact numerical 
results obtained using the algorithm of Ref. \onlinecite{ksarg.2005}. 
As expected the agreement deteriorates with decreasing the bandwidth 
$W=4|T_{\l}|$ of the reservoirs since $\S_{\rm em}^{\rm R}(\w)$ 
acquires a strong dependence on $\w$ for $\w$ in the bias window. 
The dashed lines indicate the 
steady-state value of $n_{1}$ for one-dimensional reservoirs and for WBL
reservoirs. KBE correctly approaches the one-dimensional steady-state 
in all cases whereas GKBA approaches the WBL steady-state only in the limit 
$|T_{\l}|\ra\iif$.
In the bottom left panel we consider a two-site chain 
driven out of 
equilibrium by a bias $V_{L}=-V_{R}=1$ and
again connected to one-dimensional reservoirs.\cite{mssvl.2008} In this case, however, the 
system is interacting and treated in the HF approximation. The 
chemical potential is chosen in the middle of the HOMO-LUMO gap of 
the disconnected chain with two electrons. For $T_{n}=-1$, 
$\e_{n}=0$, and Coulomb integrals $v_{11}=v_{22}=2$,  
$v_{12}=v_{21}=1$ one finds $\m=2$. The rest of the parameters are 
$T_{\l}=-1.5$ and $T=-0.5$ which, from Eq. (\ref{gamma}), implies 
$\G\simeq 0.67$ for the GKBA simulations. Even though the HOMO-LUMO 
gap $\D_{HL}=2$ is not much smaller than the bandwidth $W=4|T_{\l}|=6$ 
we still observe a satisfactory agreement for the density of 
site 1 (a similar agreement is found for site 2, not shown). The damping time as 
well as the amplitude and frequency of the transient oscillations are 
well reproduced; furthermore the GKBA steady-state value differs by less 
than 1\% from the corresponding KBE value. The accuracy of the HF-GKBA 
is not limited to the diagonal matrix elements of the density matrix. 
This is exemplified in the bottom right panel where 
we show the current flowing at the right 
interface of the four-site chain of Fig. \ref{dleads} with 
$N_{\t}=9$ transverse channels, bias $V_{L}=-V_{R}=0.8,1.2$, chemical 
potential $\m=2.26$ (chosen in the middle of the HOMO-LUMO gap 
of the disconnected chain with 4 electrons), $T_{n}=-1$, $T_{\l}=T_{\t}=-2$, $T=-0.5$, 
$\e_{n}=0$ and Coulomb integrals $v_{ii}=v=1.5$ and $v_{ij}=(v/2)/|i-j|$ 
for $i\neq j$.\cite{mssvl.2009} The GKBA and KBE currents are in excellent agreement 
except for a slight overestimation of the GKBA steady-state value at small 
bias. 

In conclusion the GKBA equation with $G^{\rm R}$ from Eq. (\ref{hfgrwbl}) and $\G$ from Eq. 
(\ref{gamma}) is a good approximation to study the HF dynamics of 
open systems provided that the embedding self-energy of the 
reservoirs has a weak frequency dependence around the chemical 
potential.

\subsubsection{Relaxation and local thermalization}

\begin{figure}[tbp]
\includegraphics[width=0.47\textwidth]{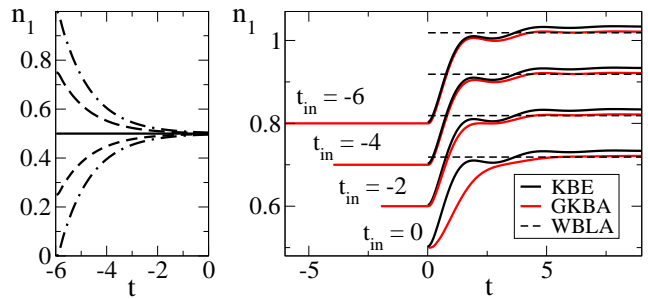}
\caption{Results for the density $n_{1}(t)$ of the one-site chain with 
$T_{\l}=-9$ and same parameters as in Fig. \ref{fig2}. 
In the left panel the system is unperturbed and 
$n_{1}(t_{\rm in})$ is varied. In the right panel $n_{1}(t_{\rm 
in})=1/2$, a bias $V_{L}=2$ is switched on at $t=0$ and $t_{\rm in}$ 
is varied. For clarity the curves with $t_{\rm in}=-2n$, $n=0,1,2,3$ 
are shifted upward by $n/10$.}
\label{fig3}
\end{figure}

For the  GKBA results of Fig. \ref{fig2} we started the propagation 
at time $t_{\rm in}<0$ with the HF density matrix of the uncontacted 
system and let $\r(t)$ thermalize in the absence of external 
fields until $t=0$  when a bias is switched on. For $t_{\rm in}$ 
sufficiently remote in the past
the density matrix attains a steady value 
$\r_{\rm eq}$ before the system is biased.
By definition $\r_{\rm eq}$ is the static solution of Eq. (\ref{eomrho}) 
with $d\r/ dt=0$; therefore if we start with $\r(t_{\rm in})=\r_{\rm eq}$
then the density matrix remains constant in the 
interval $(t_{\rm in},0)$. In the left panel of Fig. \ref{fig3} we plot the 
time-dependent density of the noninteracting one-site 
chain of Fig. \ref{fig2} for different initial values; we see that 
$n_{1}(t)=\r_{11}(t)=1/2$ for all $t<0$ if $n_{1}(t_{\rm in}=-6)=1/2$ is 
the thermalized value. It 
is  tempting to reduce the computational time (provided that 
one finds a simpler way to determine $\r_{\rm eq}$) by starting the 
propagation at  $t=0$ with  $\r(0)=\r_{\rm eq}$. 
This initial condition guarantees the local 
thermalization of all one-time observables. However, in a fully
relaxed system any two-time correlator depends on the time-difference 
only, and to achieve this relaxation a ``memory buffer'' is needed. 
Suppose that we start the propagation with 
$G^{<}(t_{\rm in},t_{\rm in})=i\r_{\rm eq}$.
Then the equal-time $G^{<}(t,t)$ remains constant but 
the $G^{<}(t,t')$ depends on $t$ and $t'$ separately. It is only  
for $t,t'$ large enough that $G^{<}(t,t')$ depends on $t-t'$.
This concept is explained in the right 
panel of Fig. \ref{fig3} where we display $n_{1}(t)$ when 
a bias $V_{L}=2$ is switched on at $t=0$.
In all cases $\r(t_{\rm in})=\r_{\rm eq}=1/2$ but the 
initial time $t_{\rm in}$ is varied. The absence of relaxation for too small 
$|t_{\rm in}|$ is evident from the strong dependence of the transient 
behavior on $t_{\rm in}$. The curves $n_{1}(t>0)$ become independent 
of $t_{\rm in}$ only for $t_{\rm in}\lesssim -4$.

The concept of relaxation, and hence of the memory buffer, 
has been illustrated in a  simple model system but its importance is 
completely general and is not limited to systems in thermal equilibrium.  
Suppose that the physical system is in some
 excited state $\r_{\rm ex}$. If we start the propagation at time $t=0$ 
with initial condition $\r(0)=\r_{\rm ex}$ then the 
transient behavior is affected by spurious relaxation processes. 
The proper way of performing GKBA simulations consists in driving the relaxed system 
toward  $\r_{\rm ex}$ with some suitable external fields.

\subsubsection{Damping}

\begin{figure}[tbp]
\begin{center}
\includegraphics[width=0.48\textwidth]{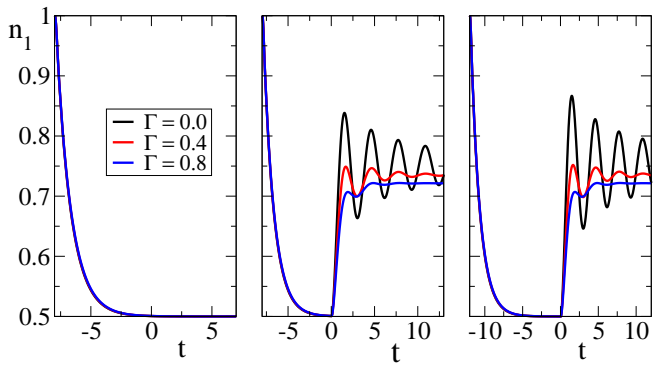}
\caption{Time dependent occupation of the one-site chain for different $\G$. Left panel: 
$t_{\rm in}=-8$ and unperturbed system. Middle panel: $t_{\rm in}=-8$ 
and bias $V_{L}=2$ switched on at 
$t=0$. Right panel: $t_{\rm in}=-12$ and bias $V_{L}=2$ switched on 
at $t=0$.}
\label{differentGG}
\end{center}
\end{figure}

For bulk systems like an electron gas the inclusion of 
damping in the propagator worsens the agreement with 
the KBE results.\cite{bsh.1999} In fact, the use of a non-Hermitian 
quasi-particle Hamiltonian 
$h_{\rm HF}-i\G/2$
in $G^{\rm R}$ is a distinctive feature of open systems. Here we 
address how sensitive the results are to 
different values of $\G$. We consider again the  noninteracting one-site 
chain of Fig. \ref{fig2} with $T_{\l}=-9$, for which Eq. (\ref{gamma}) 
yields $\G=0.8$. In all cases we set the 
initial condition  
$n_{1}(t_{\rm in})=1$.
In Fig. \ref{differentGG} (left panel) we show the relaxation 
dynamics, starting from $t_{\rm in}=-8$, of the unperturbed system 
for three different $\G$; the 
curves are essentially on top of each 
other. This may suggest that the dependence on $\G$ is  
weak. However, if we switch on a bias in the left lead $V_{L}=2$ at time 
$t=0$ (middle panel)  we appreciate a strong $\G$-dependence.
We may argue that for small $\G$  the 
relaxation time is longer and hence that the curves with 
$\G=0.0,\;0.4$ approach the curve with $\G=0.8$ by reducing $t_{\rm in}$. This  
is not the case as clearly illustrated in the right panel where 
$t_{\rm in}=-12$. The curve with $\G=0.4$ is already converged 
whereas the one with $\G=0$ is not but the trend is to 
separate further from the curve with $\G=0.8$.
The apparent weak $\G$-dependence in the 
left panel is simply due to the alignment of the on-site energy to 
the chemical potential, $\m=\e_{n}=0$. In general a proper choice of the 
quasi-particle damping is crucial for a correct description 
of the system evolution. In HF theory the damping is only due to 
embedding effects and the $\G$ of Eq. (\ref{gamma}) is the most 
accurate. 
The inclusion of correlation effects introduces an extra damping. Is it 
possible to maintain the simple form in Eq. (\ref{hfgrwbl}) for 
the quasi-particle propagator and still have good agreement with the 
KBE results? In the next Section we discuss two different 
correlated quasi-particle Hamiltonians to insert in  Eq. (\ref{hfgrwbl}).

\section{Correlated approximations to the propagator}
\label{corrpropsec}

In the interacting case the exact equation of motion for $G^{\rm R}$ reads
\be
\left[
i\frac{d}{dt}-h_{\rm HF}(t)\right]\!G^{\rm R}(t,t')=
\d(t,t')+\int \!\! d\bar{t}\,\S^{\rm R}(t,\bar{t})G^{\rm 
R}(\bar{t},t')
\label{eqgrint}
\ee
with $\S^{\rm R}=\S_{\rm em}^{\rm R}+\S_{\rm c}^{\rm R}$. If we 
approximate 
\be
\S_{\rm em}^{\rm R}(t,t')\simeq -(i/2)\G\d(t-t'),
\ee
with $\G$ from Eq. (\ref{gamma}), we find the approximate equation
\be
\left[
i\frac{d}{dt}-h^{(0)}_{qp}(t)\right]\!G^{\rm R}(t,t')=
\d(t,t')+\int \!\! d\bar{t}\,\S^{\rm R}_{\rm c}(t,\bar{t})G^{\rm 
R}(\bar{t},t')
\label{eqgrint1}
\ee
where
\be
h^{(0)}_{qp}(t)\equiv h_{\rm HF}(t)-i\G/2
\ee
is the HF quasi-particle Hamiltonian. Discarding the integral on the 
right hand side of Eq. (\ref{eqgrint1}) one finds the HF solution of 
Eq. (\ref{hfgrwbl}).  We refer to the GKBA with HF 
propagators as the GKBA0 scheme. Unfortunately the GKBA0 scheme 
performs rather poorly, see Section  \ref{results}, indicating that 
$G^{\rm R}$ has to incorporate 
correlation effects to some extent. 
Below we propose two schemes to approximate the convolution 
$\S^{\rm R}G^{\rm R}$ and
reduce Eq. (\ref{eqgrint1}) to a 
quasi-particle equation  of the form
\be
\left[
i\frac{d}{dt}-h_{qp}(t)\right]\!G^{\rm R}(t,t')=
\d(t,t').
\label{eqgrintqp}
\ee
The solution of Eq. (\ref{eqgrintqp}) is
\be
G^{\rm R}(t,t')=-i\th(t-t')\,T\,e^{-i\int_{t'}^{t}d\bar{t}\,h_{qp}(\bar{t})}
\label{qpgr}
\ee
and satisfies the group property of Eq. (\ref{group}). Therefore, if we 
are successful in this task the calculation of $G^{\rm R}$ will scale 
like $t^{2}_{\rm max}$.

\subsection{Static correlation approximation}

In open systems the correlation self-energy 
decays to zero when the separation between its time arguments 
approaches infinity. If $G^{\rm R}(\bar{t},t')\simeq G^{\rm R}(t,t')$ 
for $t-\bar{t}$ smaller than the decay time of $\S^{\rm R}_{\rm c}$ 
we can approximately write
\be
\int \!\! d\bar{t}\,\S^{\rm R}_{\rm c}(t,\bar{t})G^{\rm 
R}(\bar{t},t')\simeq \left[\int \!\! d\bar{t}\,\S^{\rm R}_{\rm 
c}(t,\bar{t})\right]G^{\rm R}(t,t').
\label{rhsint}
\ee
To evaluate the integral in the square brackets we make an adiabatic 
approximation on top of the GKBA, i.e., we replace $G^{\rm R}$ 
with the equilibrium propagator of a 
system described by the Hamiltonian $\hat{H}(t)$. Let us consider, 
for simplicity, an interaction $v_{ijmn}=\d_{in}\d_{jm}v_{ij}$. Then 
the Langreth rules\cite{svl.book,langreth} provides us with 
the following expression of the retarded 2B self-energy, see Eq. 
(\ref{2bse}),
\bea
\!\!\!\S^{\rm R}_{{\rm c},ij}(t,t')=
2\sum_{kl}v_{ik}v_{jl}[G^{\rm 
R}_{ij}(t,t')G^{<}_{lk}(t',t)G^{>}_{kl}(t,t')
\nn\\
+G^{<}_{ij}(t,t')G^{\rm A}_{lk}(t',t)G^{<}_{kl}(t,t')\!+\!
G^{<}_{ij}(t,t')G^{<}_{lk}(t',t)G^{\rm R}_{kl}(t,t')]
\nn\\
-\sum_{kl}v_{ik}v_{jl}[G^{\rm R}_{il}(t,t')G^{<}_{lk}(t',t)G^{>}_{kj}(t,t')
\nn\\
+G^{<}_{il}(t,t')G^{\rm A}_{lk}(t',t)G^{<}_{kj}(t,t')
\!+\!G^{<}_{il}(t,t')G^{<}_{lk}(t',t)G^{\rm R}_{kj}(t,t')].
\nn\\
\label{2bseret}
\eea
As $\S^{\rm R}_{\rm c}(t,t')$ vanishes for $t<t'$, the GKBA 
transforms this quantity 
into a function of $\r(t)$ and 
$G^{\rm R}(t,t')=[G^{\rm A}(t',t)]^{\dag}$. The adiabatic approximation consists in evaluating 
the GKBA form of Eq. (\ref{2bseret}) using an equilibrium propagator 
\be
\tilde{G}^{\rm 
R}(t,t-t')=\int\frac{d\w}{2\p}\frac{e^{-i\w(t-t')}}{\w-h_{qp}(t)+i\eta}
\label{tildegret}
\ee
where we use the matrix notation $1/A=A^{-1}$ for any matrix $A$.
The resulting expression, which we denote by $\tilde{\S}(t,t-t')$,
depends implicitly on $t$ through the 
dependence on $\r(t)$ and $h_{qp}(t)$ and explicitly on $t-t'$. 
If we define
\be
\tilde{\S}(t)=\int d\bar{t}\,\tilde{\S}(t,t-\bar{t})
\label{tildesw=0}
\ee
then the right hand side of Eq. (\ref{rhsint}) becomes 
$\tilde{\S}(t)G^{\rm R}(t,t')$ and Eq. (\ref{eqgrint1}) is solved by 
Eq. (\ref{qpgr}) with
\be
h_{qp}(t)=h_{\rm HF}(t)-i\G/2+\tilde{\S}(t).
\ee
In this way we generate a self-consistent equation for 
$\tilde{\S}(t)=\tilde{\S}(\r(t),h_{qp}(t))$. In practice for a given 
$\tilde{\S}(t_{n})$ at the $n$-th time step we 
determine $\r(t_{n+1})$ from Eq. (\ref{eomrho}), then calculate 
$h_{qp}(t_{n+1})=h_{\rm HF}(t_{n+1})-i\G/2+\tilde{\S}(t_{n})$, 
hence $\tilde{G}^{\rm R}(t_{n+1},t_{n+1}-t')$ and finally $\tilde{\S}(t_{n+1})$. 
Each time step can be repeated a few times to achieve convergence; in 
our experience two predictor correctors are typically enough. It is 
worth stressing that the propagator appearing in the collision 
integral is $G^{\rm R}$ and not $\tilde{G}^{\rm R}$. 
The latter is only an auxiliary quantity to calculate 
$\tilde{\S}(t)$. In the following we refer to the combination of GKBA 
with the described propagator as the  GKBA + Static Correlation 
(SC) scheme since Eq. (\ref{tildesw=0}) is the zero frequency 
value of the Fourier transform of $\tilde{\S}(t,t-\bar{t})$. 
In this  scheme the calculation of $\tilde{\S}$ for a given 
$\tilde{G}^{\rm R}$ scales like $N^{5}$ where $N$ is the number of 
basis functions in the molecular junction.

\subsection{Quasi-particle approximation}

An alternative way to introduce correlation effects in the propagator 
is again based on the adiabatic approximation but uses
the concept of quasi-particles. Let us represent operators in the
one-particle Hilbert space with a hat, e.g., $\hat{h}_{qp}$ or 
$\hat{\Sigma}^{\rm R}_{\rm c}$,  and denote by $|i\ket$ the basis ket of the 
molecular junction so that $\bra i|\hat{\Sigma}^{\rm R}_{\rm 
c}|j\ket=\hat{\Sigma}^{\rm R}_{{\rm c},ij}$, etc.. For an isolated 
molecule in equilibrium the quasi-particle equation reads
\be
[\hat{h}_{\rm HF}+\hat{\Sigma}^{\rm R}_{\rm c}(\e)]|\vf\ket=\e|\vf\ket
\label{qpeq}
\ee
where $\hat{\Sigma}^{\rm R}_{\rm c}(\e)$ is the Fourier transform of 
the equilibrium self-energy. To lowest order in 
$\Sigma^{\rm R}_{\rm c}$ this equation implies that the correction to 
the HF energies $\e_{{\rm HF},n}$ is 
\be
\e_{qp,n}=\e_{{\rm HF},n}+\bra\vf_{n}|\hat{\Sigma}^{\rm R}_{\rm 
c}(\e_{{\rm HF},n})|\vf_{n}\ket
\label{qpeq1st}
\ee
where $|\vf_{n}\ket$ is the eigenket of $\hat{h}_{\rm HF}$ with eigenvalue 
$\e_{{\rm HF},n}$. Equation (\ref{qpeq1st}) suggests to construct a quasi-particle 
Hamiltonian in the following manner. We evaluate again the GKBA form of Eq. 
(\ref{2bseret}) with the propagator of Eq. (\ref{tildegret}) and then 
calculate
\be
\tilde{\S}(t,\w)=\int dt \,e^{i\w(t-t')}\tilde{\S}(t,t-t').
\ee
From this quantity we construct the one-particle operator
$\hat{\tilde{\S}}(t,\w)=\sum_{ij}|i\ket \tilde{\S}_{ij}(t,\w)\bra 
j|$ and subsequently the diagonal self-energy operator in the HF basis
\be
\hat{\tilde{\S}}(t)=\sum_{n}| \vf_{n}\ket
\bra\vf_{n}|\hat{\tilde{\Sigma}}(t,\e_{{\rm 
HF},n})|\vf_{n}\ket\bra\vf_{n}|.
\ee
Imposing now that 
$\hat{h}_{qp}(t)=\hat{h}^{(0)}_{qp}(t)+\hat{\tilde{\S}}(t)$ we get
a self-consistent equation for $\tilde{\S}(t)$. We refer to this 
procedure as the GKBA + Quasi-Particle (QP) scheme. As the Fourier 
transform of $\tilde{\S}(t,t-t')$ has to be evaluated in $N$ different energies
the calculation of $\tilde{\S}(t)$ in the GKBA+QP scheme scales like $N^{6}$.

\section{Results}
\label{results}

In this Section we study the nonequilibrium {\em correlated} dynamics of 
the chain junction of Fig. \ref{dleads} and of a model photovoltaic 
junction. We calculate local occupations, currents, and spectral 
functions using different GKBA schemes, and benchmark the results 
against full KBE simulations. A clear-cut scenario will emerge in which  
GKBA+SC is the most reliable scheme while
all other schemes suffer from some deficiencies.

\subsection{Chain junction}
\label{reschain}
\begin{figure}[tbp]
    \begin{center}
	\includegraphics[width=0.48\textwidth]{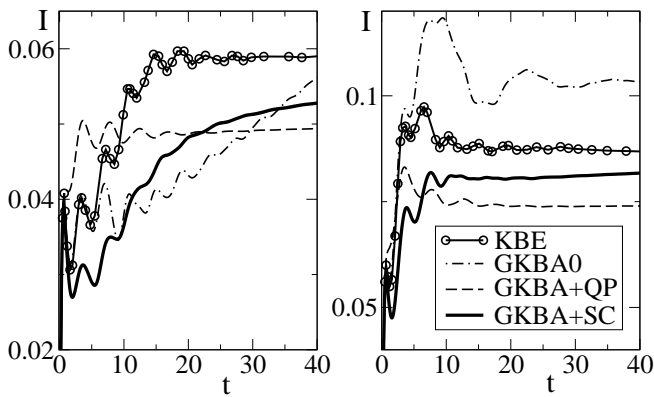}
	\caption{Time-dependent current at the right interface of the four-site 
	junction with $V_{L}-V_{R}=1.6$ (left 
	panel) and 2.4 (right panel); same parameters as in 
	Fig. \ref{fig3}. }
	\label{2Bchain}
    \end{center}
\end{figure}
Nonequilibrium correlation effects change drastically the HF 
picture of quantum transport. The applied bias causes an 
enhancement of quasi-particle scatterings and consequently a substantial 
broadening of the spectral peaks.\cite{mssvl.2009,t.2009} The 2B 
steady current is 
larger (smaller) than the HF steady current 
at bias  smaller (larger)  than 
the HF HOMO-LUMO gap, see bottom-right panel 
of Fig. \ref{fig3}. In Fig. \ref{2Bchain} we compare the current at 
small (left panel) and large (right panel) bias using  
KBE and different GKBA schemes. Even though the 
correlation-induced enhancement (at small bias $V_{L}-V_{R}=1.6$ the HF 
steady current is $\sim 0.023$) 
and suppression (at large bias $V_{L}-V_{R}=2.4$ the HF 
steady current is $\sim 0.11$) of the steady current relative to the HF values is 
qualitatively captured by all GKBA schemes, quantitative 
differences emerge. GKBA0 is rather close to KBE during the 
initial transient but considerably overestimates the steady state. 
GKBA+QP corrects too much this deficiency and the steady current is 
appreciably underestimated. Furthermore the transient behavior worsens: the first 
peak is absent and the current saturates too fast. This is due to a general problem of the GKBA+QP 
scheme. The equilibrium $\tilde{\S}$ 
is too large or, equivalently, equilibrium correlations are 
overestimated. GKBA+SC gives an overall improvement. The transient 
current reproduces several KBE features (oscillation frequency and relative 
hight of the peaks) and the steady current is very close 
to the KBE value. 

\begin{figure}[tbp]
    \begin{center}
	\includegraphics[width=0.48\textwidth]{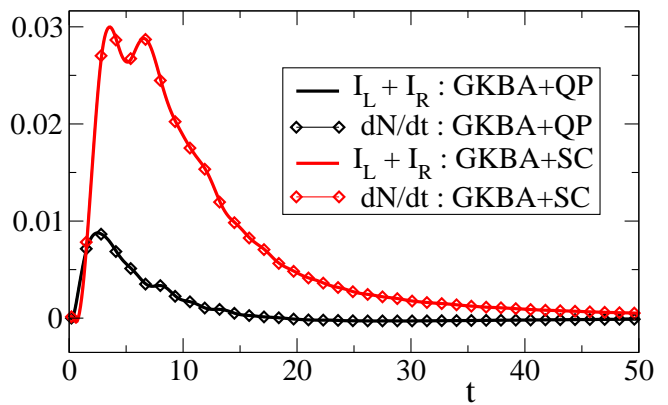}
	\caption{Numerical evidence of the fulfillment of the
	continuity equation in the GKBA+QP and GKBA+SC schemes. }
	\label{continuity}
    \end{center}
\end{figure}
By construction the GKBA schemes guarantee the satisfaction of the 
continuity equation. The rate of change of the 
total number of electrons in the nanostructure, $dN/dt$, is equal to 
the sum of the currents flowing through the left and right interface, 
$I_{L}+I_{R}$. In Fig.  \ref{continuity} we show that this analytic 
property is numerically confirmed with high accuracy in the GKBA+QP 
and GKBA+SC schemes.

\subsection{Photovoltaic junction}

\begin{figure}[b]
\begin{center}
\includegraphics[width=0.45\textwidth]{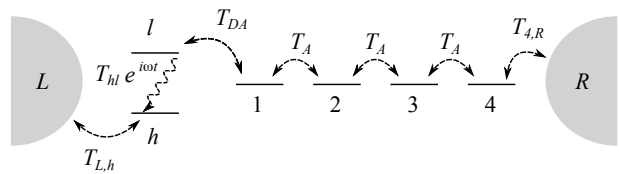}
\caption{Schematic illustration of the photovoltaic junction 
described in the main text.}
\label{fig:junction}
\end{center}
\end{figure}

We consider a more complicated open system with the features of a 
photovoltaic molecular junction.
Inspired by a paper by Li et al.\cite{lnr.2012} we model the 
junction as a donor--acceptor complex connected to a left and right 
electrodes (reservoirs), see Fig. \ref{fig:junction}. The donor is 
described by HOMO ($h$) and LUMO ($l$) levels and the LUMO 
is connected to a chain of 
four acceptor sites each described by a single localized orbital. 
These orbitals are mixed by the acceptor Hamiltonian and form 
two valence and two conduction levels. The junction is connected to 
the left electrode throught the HOMO with tunneling amplitude $T_{L,h}$ 
and to the right electrode through 
the rightmost acceptor site with tunneling amplitude $T_{4,R}$.     
The explicit form of the Hamiltonian of the donor--acceptor complex is
\bea
\hat{H}_{J}&=&\e_{h}\hat{n}_{h}+\e_{l}\hat{n}_{l}+\e_A
\sum_{a=1}^{4}\hat{n}_a
\nn\\
&+&T_{DA}\sum_{\s}\left(
\hat{d}_{l\s}^{\dag}\hat{d}_{1\s}+\hat{d}_{1\s}^{\dag}\hat{d}_{l\s} \right)
\nn\\
&+&T_A\sum_{a=1}^{3}\sum_{\s}\left(
\hat{d}_{a\s}^{\dag}\hat{d}_{a+1\s}+\hat{d}_{a+1\s}^{\dag}\hat{d}_{a\s}
\right)
\nn\\
&+&U_{DA}\left(\hat{n}_{h}+\hat{n}_{l}-2\right)
\sum_{a=1}^{4}\frac{\hat{n}_a-1}{a},
\eea
where $\hat{n}_{x}=\sum_{\s}\hat{d}^{\dag}_{x\s}\hat{d}_{x\s}$ is the occupation 
operator for $x=h,l,a$. The interaction between the excess charges of 
the donor and acceptor chain implicitly fixes the condition of charge 
neutrality. For one-dimensional reservoirs with longitudinal hopping 
integral $T_{\l}=-9$, 
tunneling amplitudes
$T_{L,h}=T_{4,R}=-0.3$, donor levels $\e_{h}=-2.92$, $\e_{l}=-0.92$, 
acceptor levels $\e_{A}=-2.08$, donor-acceptor hopping $T_{DA}=-0.1$, 
intra-acceptor hopping $T_{A}=-0.2$ and interaction $U_{DA}=0.5$,
the chemical potential $\m=0.04$ is in the middle of the HF gap 
between the valence and conduction acceptor levels.
The equilibrium system has HOMO and LUMO occupations 2 and 0 respectively 
and the two valence levels of the acceptor chain completely filled.
The photovoltaic junction is driven out of equilibrium by  
irradiation with monochromatic light. For simplicity we assume that the light 
couples only to the donor 
dipole moment and hence 
\be
\hat{H}_{light}(t)=s(t)T_{hl}\sum_{\s}\left(
e^{i\w t}\hat{d}^{\dag}_{h\s}\hat{d}_{l\s}
+e^{-i\w t}\hat{d}^{\dag}_{l\s}\hat{d}_{h\s}\right),
\label{eq:laserperturbationtHL}
\ee
where $s(t)$ is a switching function. We consider $T_{hl}=0.3$, $\w=2=|\e_{h}-\e_{l}|$ 
and study a pulse, $s(t)=1$ for $0<t<\p/T_{hl}$ and zero otherwise, as 
well as continuous radiation, $s(t)=1$ for $t>0$ and $s(t)=0$ for 
$t<0$.

In order to apply many body perturbation theory we 
rewrite $\hat{H}_{J}$ in the form of Eq. (\ref{hamnano}). The 
one-particle Hamiltonian $h_{ij}$ with $i,j=h,l,a$ then reads 
\be
h = \begin{pmatrix}
  \tilde{\e}_{h} & 0 & 0 & 0 & 0 & 0  \\
  0 & \tilde{\e}_{l} & T_{DA} & 0 & 0 & 0  \\
  0 & T_{DA} & \tilde{\e}_{A\,1} & T_{A} & 0 & 0  \\
  0 & 0 & T_A & \tilde{\e}_{A\,2} & T_A & 0  \\
  0 & 0 & 0 & T_A & \tilde{\e}_{A\,3} & T_A  \\
  0 & 0 & 0 & 0 & T_A &\tilde{\e}_{A\,4}
\end{pmatrix}
\ee
with $\tilde{\e}_{h}=\e_{h}+\frac{25}{12}U_{DA}$, 
$\tilde{\e}_{l}=\e_{l}+\frac{25}{12}\,U_{DA}$ and 
$\tilde{\e}_{A\,a}=\e_{A}+\frac{25}{6}\,U_{DA}-2\,U_{DA}/a$. 
For the Coulomb integrals we find $v_{ijmn}=\d_{in}\d_{jm}v_{ij}$ with 
\be
v =  U_{DA}
 \begin{pmatrix}
  0 & 0 & 1 & 1/2 & 1/3 & 1/4  \\
  0 & 0 & 1 & 1/2 & 1/3 & 1/4  \\
  1 & 1 & 0 & 0 & 0 & 0  \\
  1/2 & 1/2 & 0 & 0 & 0 & 0  \\
   1/3 &  1/3 & 0 & 0 & 0 & 0  \\
   1/4 &  1/4 & 0 & 0 & 0 & 0 
 \end{pmatrix}.
\ee

\begin{figure}[tbp]
    \begin{center}
	\includegraphics[width=0.48\textwidth]{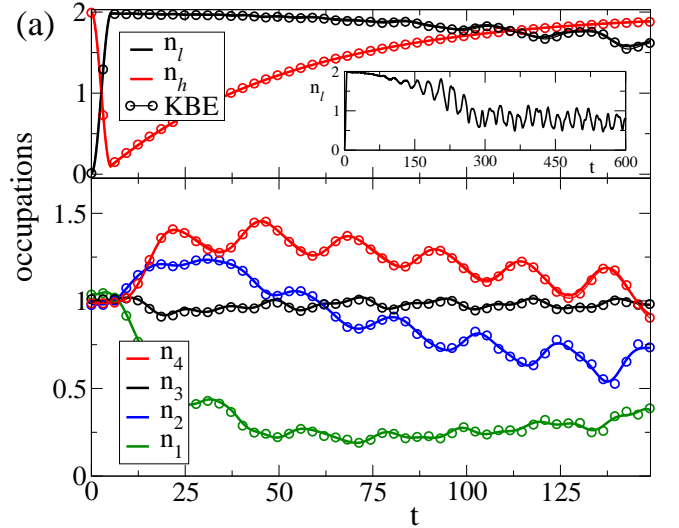}
	\caption{Time-dependent occupations in the HF approximation using  
	GKBA (solid) and KBE (circles). The junction is perturbed by 
	a monochromatic pulse.}
	\label{solar_pulse_HF}
    \end{center}
\end{figure}
Let us start with the mean-field analysis of the light pulse. The duration 
$\p/T_{hl}$ has been chosen to get a population inversion of the HOMO 
and LUMO levels. In Fig. 
\ref{solar_pulse_HF} we show the HF occupations of the donor (top 
panel) and 
acceptor (bottom panel) levels in GKBA and KBE. The impressive agreement is due to 
the fact that for $T_{\l}=-9$ the WBL approximation is extremely good.
The depletion of charge on the first acceptor site (A1) is a consequence 
of the repulsive interaction $U_{DA}$. During the pulse the HOMO level is 
partially refilled by the left reservoir and the total charge on the 
donor overcomes 2. This excess charge is instantaneously felt by A1 
which starts expelling electrons at a rate 
larger than the tunneling rate from LUMO to A1. We also observe that the charge transfer 
between LUMO and A1 is not effective. The inset shows the LUMO 
occupation on a longer time scale. Electrons remain trapped and slosh 
around along the junction. In fact, in HF no steady-state is reached. 
The occurrence of self-sustained charge oscillations in mean-field 
treatments has been observed in similar contexts\cite{mtksvl.2012,kusskvlg.2012} and is most 
likely an artifact of the approximation. As we shall see,
the correlated KBE results are very different. Therefore the collision integral and the 
correlated propagator of the GKBA  approach have to correct the HF theory in a 
substantial manner.

\begin{figure}[tbp]
    \begin{center}
	\includegraphics[width=0.48\textwidth]{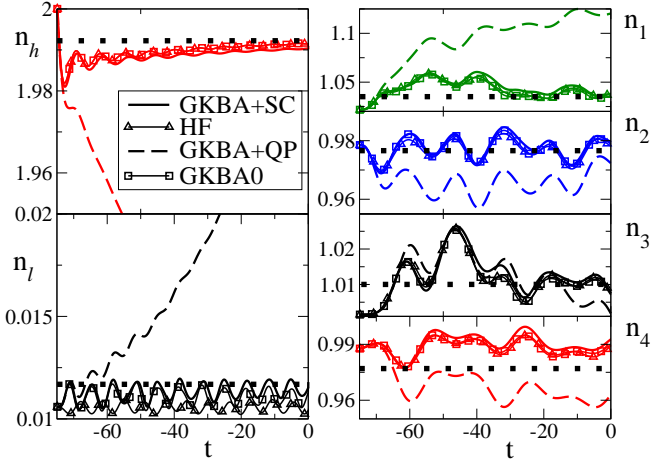}
	\caption{Thermalization of the occupations in the correlated 
	case. The correlated KBE  value is represented by a dotted 
	horizontal line.}
	\label{thermalization}
    \end{center}
\end{figure}
With the inclusion of correlations a deficiency of the GKBA+QP scheme 
emerges already during the thermalization 
process. In Fig. \ref{thermalization} the donor and 
acceptor occupations are propagated within different schemes in the 
absence of external fields  
using the HF value of the uncontacted system  and 
$\tilde{\S}(t_{\rm in})=0$ as initial conditions. 
Both GKBA0 and GKBA+SC thermalize, similarly to HF, to values very 
close to the equilibrium values of the correlated (2B) KBE approach 
(dotted horizontal line). In fact, in  
KBE  the HF and 2B equilibrium occupations are essentially 
the same since the correlation self-energy, except for a slight 
renormalization of the quasi-particle energies (image charge effect), 
does not affect the width of the spectral peaks. For the GKBA to 
reproduce the KBE thermalized values the imaginary part of 
$\tilde{\S}$ has to be small, and this is not the case in GKBA+QP. 
Here $\Im[\tilde{\S}_{ll}(t)]$ and $\Im[\tilde{\S}_{hh}(t)]$ tend to 
increase thus broadening the HOMO and LUMO spectral peaks.  
Hence the HOMO looses charge whereas the LUMO acquires charge and the donor 
polarizability increases. This makes the first bubble diagram of 
the 2B self-energy larger, and therefore the HOMO and LUMO spectral 
peaks more broadened. In a separate calculation (not shown) we 
simulated the GKBA+QP thermalization and found that the 
thermalization process is extremely slow, $t_{\rm in}\lesssim -1000$,
and that the thermalized 
value of, e.g., the LUMO occupation is $\sim 0.7$, well above the KBE 
result.

\begin{figure}[tbp]
    \begin{center}
	\includegraphics[width=0.48\textwidth]{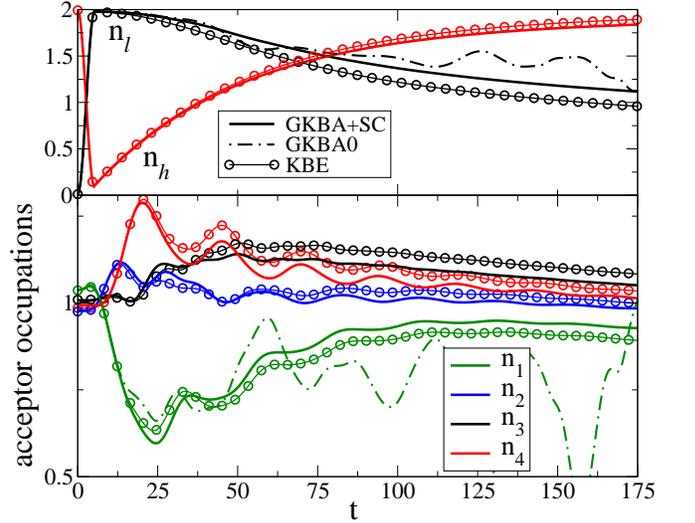}
	\caption{Time-dependent occupations after a pulse in KBE and 
	GKBA+SC. For 
	$n_{l}$ and $n_{1}$ we also show the results of the GKBA0 
	scheme (dotted-dashed).}
	\label{densities_pulse}
    \end{center}
\end{figure}
We are now ready to show the correlated results in the case of a 
light pulse. The KBE occupations are shown in Fig. \ref{densities_pulse}  
and are considerably different from the HF occupations of Fig. 
\ref{solar_pulse_HF}. The GKBA+SC scheme is in fairly good agreement 
with KBE {\em for all} occupations. 
To illustrate the crucial role 
played by our correlated propagator we also display the 
LUMO and A1 occupations in the GKBA0 scheme (dotted-dashed line). 
Even though the initial transient is acceptable the GKBA0 occupations 
become soon inaccurate. 
Therefore the evaluation of the GKBA collision integral with HF 
propagator performs rather poorly in open systems. The  GKBA+SC 
scheme has the merit of working both in and out of equilibrium.

\begin{figure}[tbp]
    \begin{center}
	\includegraphics[width=0.48\textwidth]{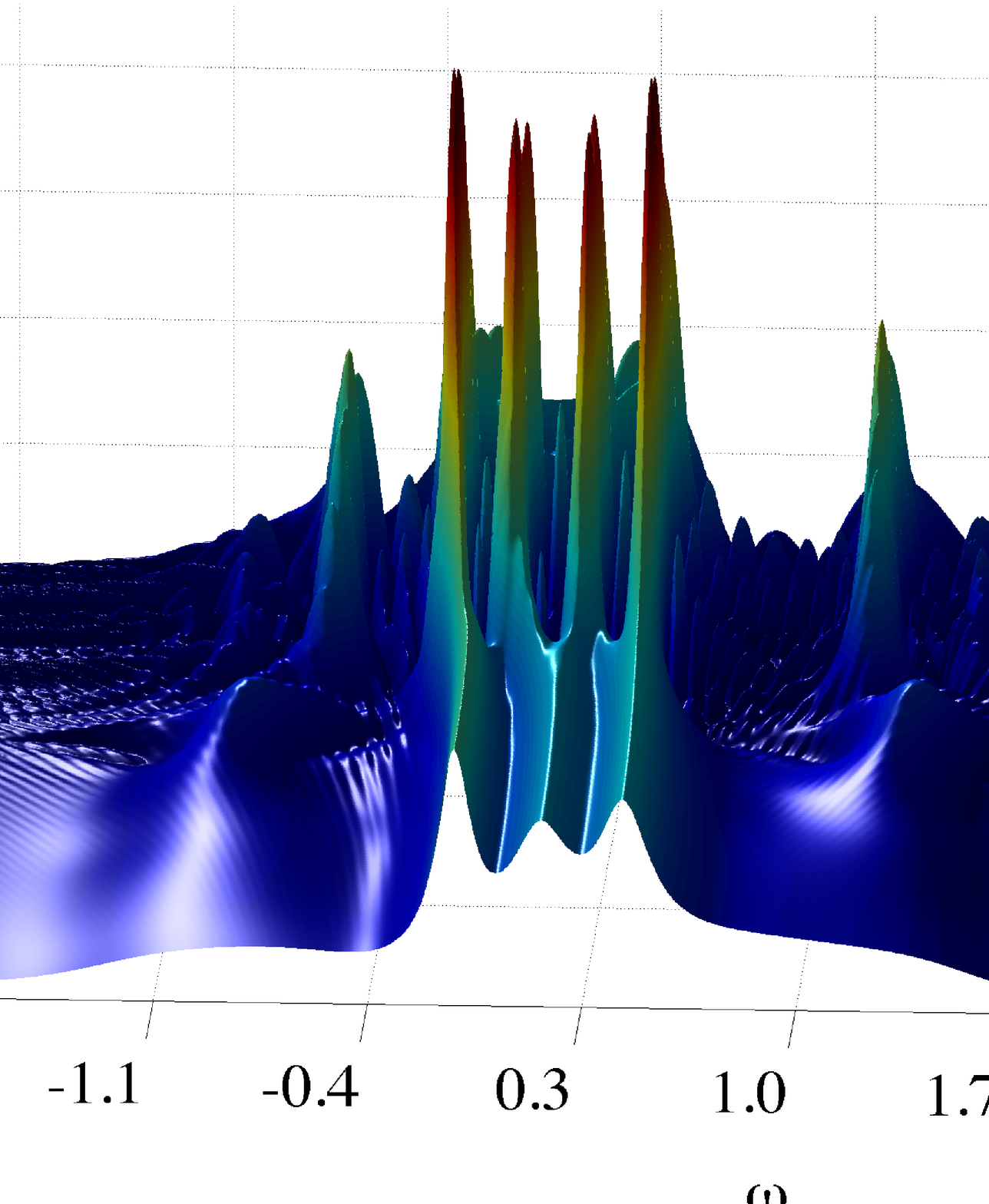}
	\includegraphics[width=0.46\textwidth]{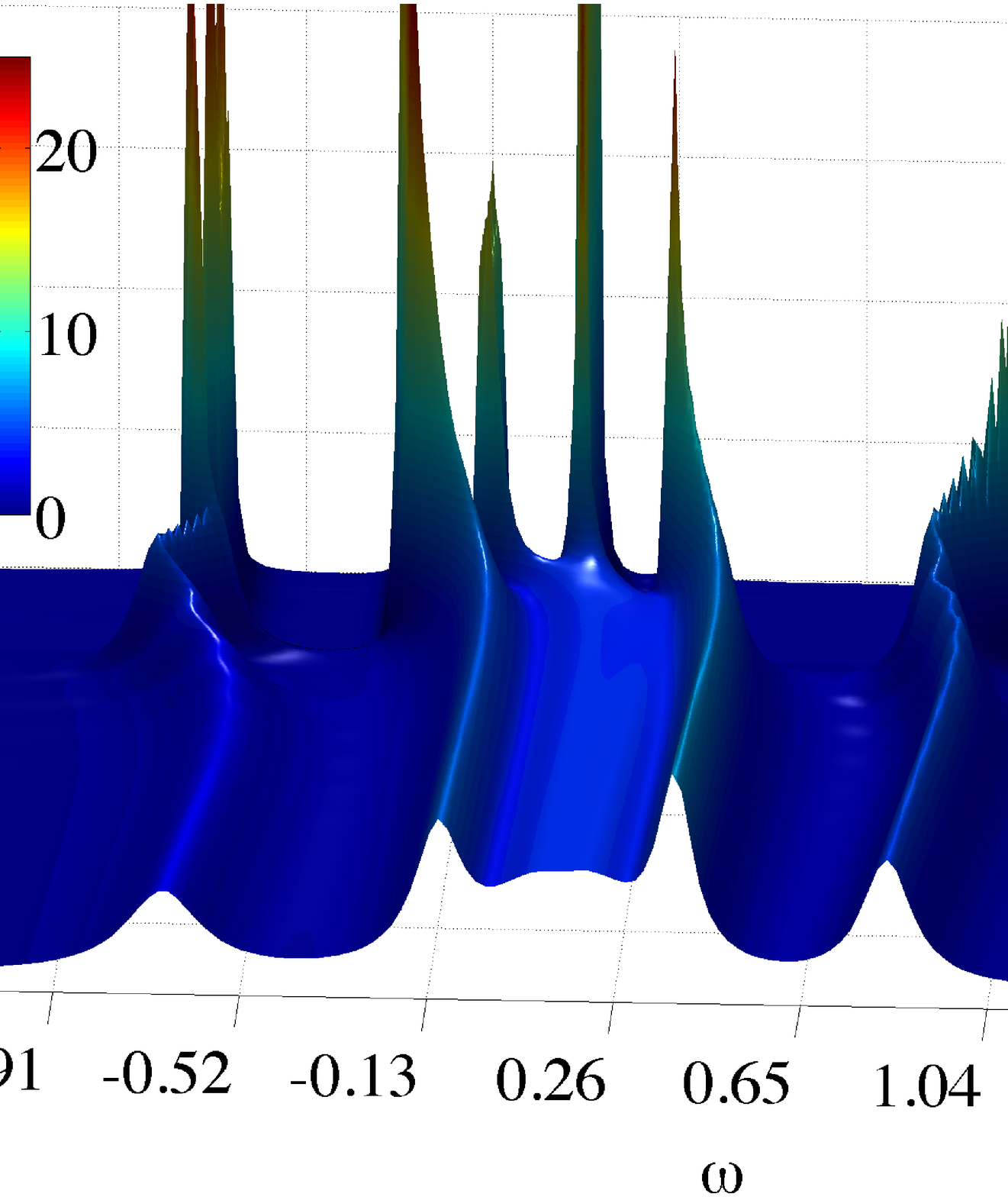}
	\caption{Time-dependent KBE (top panel) and GKBA+SC (bottom 
	panel) spectral functions for the photovoltaic junction 
	subject to a light pulse. The light pulse is switched on at 
	time $t=40$ for the KBE and $t=75$ for the GKBA+SC. The parameters are the same as in 
	Fig. \ref{densities_pulse}.}
	\label{tdspectra}
    \end{center}
\end{figure}
Like the only goal of TDDFT is to reproduce the density of an interacting 
system, so the only goal of the GKBA is to reproduce the density matrix of an 
interacting system. The TDDFT or GKBA spectral 
function $A(t,t')=i[G^{\rm R}(t,t')-G^{\rm A}(t,t')]$ can be very 
different from the true one. This is, however, not always the case. 
In Fig. \ref{tdspectra} we show the time evolution of the KBE and GKBA+SC 
total spectral function defined according to
\be
A(T,\w)=-2\Im\int d\t e^{i\w\t}\Tr\left[G^{\rm 
R}(T+\frac{\t}{2},T-\frac{\t}{2})\right],
\ee
where $T=(t+t')/2$ is the center of mass time
and $\t=t-t'$ is the relative time.
Remarkably the two spectral functions have several common features.
The most important one is the broadening of the spectral peaks after the 
pulse and the long elapsing time to relax back to the equilibrium state.
Another common feature is the drift of the acceptor peaks toward 
higher energy and the merging of the two middle peaks of the acceptor 
chain. In GKBA0 the spectral peaks are sharp at all times whereas 
in GKBA+QP they are broadened at all times (not shown).

\begin{figure}[tbp]
    \begin{center}
	\includegraphics[width=0.48\textwidth]{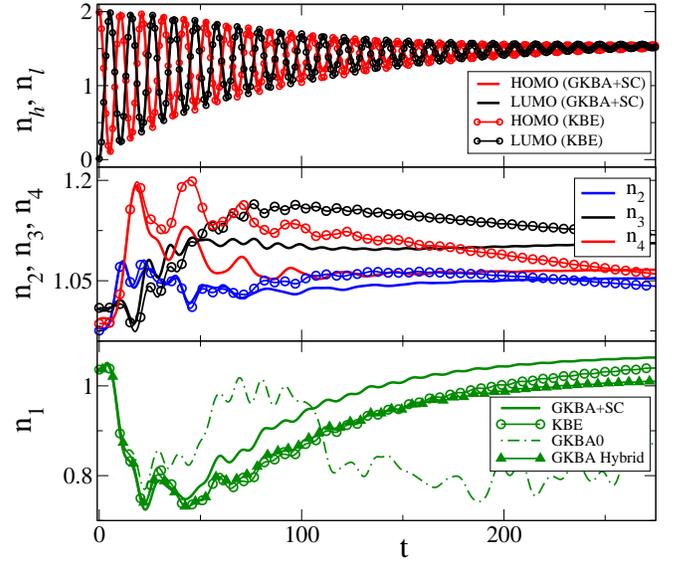}
	\caption{Time-dependent occupations in the presence of 
	continuous radiation in KBE and GKBA+SC. For $n_{1}$ we  also
	show the results of the GKBA0 scheme as well as of the hybrid scheme 
	(thermalization with GKBA+SC, positive-time propagation 
	 with GKBA+QP).}
	\label{densities_radiation}
    \end{center}
\end{figure}
To end our discussion on the performance of GKBA in open systems we 
consider in Fig. \ref{densities_radiation} the occupations for the 
continuous radiation.  Here GKBA+SC is not as accurate as in the 
case of the light pulse. However, the agreement with  KBE remains 
satisfactory. The HOMO and LUMO occupations are 
essentially indistinguishable from the KBE values (top panel). The occupations  of the 
acceptor sites next to the right electrode (A3 and A4)  are slightly 
underestimated in GKBA+SC but the overall trend (transient oscillations and 
steady-state value)  are correctly reproduced  (middle panel). A more 
quantitative agreement is observed for the acceptor sites 
next to the donor (A1 and A2). For the A1 occupation 
we also show the GKBA0 occupation (bottom panel) and we note again that 
after a short time the result deviates considerably from the KBE 
result. 

Is there any possibility of improving over the  GKBA+SC scheme  
using a different $\tilde{\S}$, or the only way is to go beyond the GKBA? 
In the bottom panel of Fig. \ref{densities_radiation} we display the 
A1 occupations for a hybrid scheme in which $\tilde{\S}$ is 
calculated from GKBA+SC at negative times (thermalization) 
and from GKBA+QP at positive times. The improvement up to 
times $t\sim 200$ is impressive and extend to {\em all} acceptor 
occupations (not shown). Instead, for times $t>200$ the KBE results 
are closer to those of the  GKBA+SC scheme. More generally for 
$t\lesssim 200$ 
we observed 
that the hybrid scheme performs better than  GKBA+SC for 
$\w\sim \e_{h}-\e_{l}$ (large current in the junction) and worse 
otherwise (small current in the junction). The purpose of this 
investigation is to provide numerical evidence of the existence 
of  a $\tilde{\S}$ for accurate GKBA simulations, and hence the 
possibility of improving  the GKBA+SC scheme without increasing the 
computational cost.

\section{Summary and outlook}

We demonstrated that time-dependent NEGF simulations of 
molecular junctions (and more generally open quantum systems) 
can be considerably speeded up. Different GKBA-based 
schemes have been proposed and subsequently 
benchmarked against full KBE calculations. The GKBA+SC scheme turned out to be 
the most accurate both in and out of equilibrium, while still offering a 
significant computational gain (for the longest propagation 
($t_{\rm max}=300$) of the photovoltaic junction the CPU time is $\sim$ 10 
minutes in GKBA+SC and $\sim$ 20 hours in  KBE). We also showed that  the  GKBA+SC 
scheme can, in principle, be further improved  without rising the 
computational price. 

All calculations have been performed within the 2B approximation for 
the correlation self-energy but the GKBA+SC scheme is completely 
general and not limited to this special case. Clearly, in large nanostructures 
screening is important and the interaction should be treated, at 
least,  within the GW approximation. Another urgent extension of the  GKBA+SC scheme 
is the inclusion of  
the interaction between electrons and nuclei. This can be done either at the level of the Ehrenfest 
approximation\cite{vsa.2006,rozzi.2013} or by adding 
diagrams with electron-phonon vertices to the correlation 
self-energy.\cite{m.2013} 

The GKBA+SC scheme, its extensions and refinements
can be implemented in {\em ab initio} molecular 
codes\cite{yambo} to perform first principle time-dependent simulations of
open nanostructures. Foreseeable applications are, e.g., in the  
field of molecular photovoltaics and molecular electronics. 
Here there is much interest in developing 
efficient quantum simulation methods for an accurate description of 
the electron-hole formation, recombination and separation as well as of charge 
transfer and possibly ionic reorganization or isomerization. 
In molecular photovoltaics ab-initio studies have focused 
on the optical spectra using linear response TDDFT\cite{afs.2008}
or the Bethe-Salpeter equation.\cite{bao.2011} Real-time 
simulations remain, however, the most powerful tool to resolve the 
different competing processes up to the ps time scale. 
State-of-the-art simulations treat the contacts as 
finite-size clusters while taking into account the full atomistic 
structure either semi-empirically\cite{rb.2003}
or fully ab initio.\cite{dsp.2005,dp.2007} However, these 
studies suffer from spurious boundary effects like 
the formation of artificial electric fields and reflection of charge 
after a few tens of fs. There are no such limitations in 
the GKBA for open systems as the electrodes are described in a 
virtually exact way through the embedding self-energy. Furthermore 
the effects of the Coulomb interaction can be systematically included through the 
diagrammatic expansion of the correlation self-energy.
The encouraging results presented in this work 
should foster advances in the development of a NEGF approach to 
ultrafast  processes at the nanoscale. 

\subsection*{Acknowledgements}

E. Perfetto and G. Stefanucci acknowledge funding by MIUR FIRB Grant
No. RBFR12SW0J. R. van Leeuwen and A.-M. Uimonen 
acknowledge support from the Academy of Finland as well as the CSC IT 
center for providing resources for scientific computing.


\begin{thebibliography}{10}
\bibitem{nota3}
The GKBA is an ansatz for the lesser/greater Green's function.
At present there exists no extension of the GKBA to the left/right 
Green's functions\cite{svl.book} for the inclusion of initial correlations with the addition 
of a vertical track (imaginary Matsubara axis). Therefore, relaxation 
can only be achieved by propagating the unperturbed systems from 
$t_{\rm in}<0$ to time $t=0$.
\end{thebibliography}
\end{document}